\documentclass{ws-ijmpcs}
\usepackage{epstopdf}
\begin{document}

\markboth{R.N. Bhatt, Sonika Johri}
{``Rare'' Fluctuation Effects in the Anderson Model of Localization}

%
\catchline{}{}{}{}{}
%

\title{``RARE'' FLUCTUATION EFFECTS IN  THE ANDERSON MODEL OF LOCALIZATION}

\author{R.N. BHATT}

\address{Department of Electrical Engineering and Princeton Center for Theoretical Science, Princeton University\\
Princeton, NJ 08544, USA \\
ravin@princeton.edu}

\author{S. JOHRI}

\address{Department of Electrical Engineering, Princeton University\\
Princeton, NJ 08544, USA\\
sjohri@princeton.edu}

\maketitle

\begin{history}
\received{19 March 2012}
\revised{Day Month Year}
\end{history}

\begin{abstract}
We discuss the role of rare fluctuation effects in quantum condensed matter systems. In particular, we present recent numerical results of the effect of resonant states in Anderson's original model of electron localization. We find that such resonances give rise to anomalous behavior of eigenstates not just far in the Lifshitz tail, but rather for a substantial fraction of eigenstates, especially for intermediate disorder. The anomalous behavior includes non-analyticity in various properties as a characteristic. The effect of dimensionality on the singularity, which is present in all dimensions, is described, and the behavior for bounded and unbounded disorder is contrasted.

\end{abstract}

\ccode{PACS numbers: 11.25.Hf, 123.1K}

\section{Introduction: A Brief History of Large Disorder and Rare Fluctuations}	

Rare fluctuation phenomena received little attention in the early days of Solid State Physics. Ignoring them seems to have been the norm –-- what effects could such low probability phenomena possibly have, especially for a thermodynamically large system? On the contrary, the phenomenal success of Bloch's theorem and its application to wave phenomena in solids (e.g. electron energy bands, phonon dispersions), led to a kind of ``Blochitis'' among the practitioners of the field. Materials not having proper translational symmetry (i.e. disordered materials like amorphous solids, glasses, liquids) were addressed using perturbative techniques starting from the uniform system, usually a crystal with an appropriate coordination number, and then averaging the variables representing the disorder.  This led to the growth of various mean-field like schemes for treating disorder, like the Coherent Potential Approximation, the averaged T-matrix approximation, the Effective Medium Approximation etc. 

However, none of these methods quite captured effects that were primarily present because of disorder, as exemplified in the seminal paper by Anderson\cite{anderson} in 1958 entitled ``Absence of Diffusion in Certain Random Lattices''. Anderson showed, for the first time, the existence of localized states in a disordered one-electron system, localized around disorder-specific locations in the system, in contrast to the Bloch waves in crystalline materials, which extended in a periodic fashion over the whole macroscopic system at all allowed energies. In our three dimensional world, localized states existed for moderate to large disorder. Further, as emphasized by Mott\cite{mott}, these localized states were separated in energy from extended states; the separatrix was called the mobility edge; in recent years, it is also referred to as the critical energy of the localization-delocalization transition, $E_C$. Thus, with increasing disorder, localized states appeared first in regions of low density of states (“band tails”), and gradually the mobility edge(s) moved towards the center of the band, until at a critical disorder, all states became localized and no extended Bloch-like states were left. (In reality, extended states for moderate disorder also exhibit large-scale fluctuations that are disorder specific, and are known in the field as mesoscopic fluctuations). 

Soon after Anderson's paper, Mott and Twose\cite{motttwose} showed that all electronic states were localized in one dimension for arbitrarily weak disorder (and hence Bloch's ideas broke down immediately upon introduction of disorder, at least as far as electron transport was concerned). It took almost two decades, using renormalization group (RG) ideas, to establish that in two dimensions\cite{abrahams} all states were again localized with arbitrary small disorder in the case of pure potential scattering, like the one considered by Anderson in his 1958 paper. In hindsight, given the situation in lower dimensions, it seems lucky that we live in a three-dimensional world, where many of the properties of crystalline materials we love and use are robust, at least to small amounts of imperfections.

Following Anderson's stylized model, Lifshitz\cite{lifshitz} considered a somewhat more realistic model for problem of impurity bands in doped semiconductors, in which positional randomness of the dopants was the dominant source of disorder. Here the dopant atom, often less than one part in a million, dominates all the characteristics of the material that give it the unique properties that led to the semiconductor revolution. It is ironic to note that while solid state physicists studied doped semiconductors quite extensively since the 1950s, nobody seems to have thought of that (i.e., physics of dopants) as a rare fluctuation effect! 

In modern parlance, Lifshitz considered the problem with purely off-diagonal (hopping) disorder, whereas Anderson had dealt with purely diagonal (onsite energy) disorder. Many short and intermediate distance properties are better captured by Lifshitz's model of the impurity band, especially at low dopant densities. In his analysis, Lifshitz found that band tails were formed as a result of resonance between pairs of sites that happened to be much closer than average, a rare fluctuation effect. An analogous study of band tails in the Anderson model can also be performed, leading to “Lifshitz tails” on rare configurations of clusters consisting of resonant sites, with many analytic results being arrived at near the band edge\cite{lifshitz}\cdash\cite{kramer}.  Soon thereafter, band tails were studied in the high-density limit by Halperin and Lax using minimum counting methods\cite{halperin}.

The next major appearance of rare fluctuation effects in condensed matter appears in two studies of many-body spin systems, by McCoy and collaborators\cite{mccoywu}\cdash\cite{mccoy} and by Griffiths\cite{griffiths}. These studies, which appeared around 1968-69, dealt with classical, Ising models of magnetism with disorder. The former was a study of a two-dimensional square Ising ferromagnet with uniform vertical bonds and random bonds in the horizontal direction, which were perfectly correlated in the vertical dimension - it thus appeared as a contrived model in the classical context.  However, the results obtained were quite bizarre - it was found that the magnetic susceptibility diverged due to rare configurations of strong bonds in a finite region in the paramagnetic phase above the true thermodynamic phase transition temperature. Griffiths, on the other hand, was studying an Ising ferromagnet in any dimension with randomly diluted bonds. He showed that the existence of rare configuration of clusters with fewer than average missing bonds over finite length scales led to essential singularities in the thermodynamics of the system in the paramagnetic phase, between the transition temperature of the pure, undiluted system, and the actual transition temperature of the diluted system. While these weak singularities in the thermodynamics were experimentally undetectable, it was shown several years later\cite{dhar} that there was a concomitant slow, non-exponential decay in the long time dynamics of such systems.  Such effects were also claimed\cite{randeria} to affect the long time dynamics of Ising spin glasses (where the bonds are ferromagnetic and antiferromagnetic at random with equal probability); while non-exponential relaxation\cite{ogielski}\cdash\cite{svedlindh} has been a hallmark of spin glasses, whether the long time behavior is dominated by rare configurations remains an open issue at present.

Roughly a decade after Griffiths and McCoy, the problem of randomly positioned donors surfaced again\cite{bhattrice}, this time in the problem of the magnetic ground state of n-doped semiconductors, which could be modeled as a quantum spin-1/2 Heisenberg antiferromagnet with an extremely high degree of disorder (large randomness). This was addressed using an RG scheme that made use of large disorder, leading to bond distributions that were wide on a logarithmic scale\cite{bhattlee}. A one-dimensional analog of the same problem allowed an analytic treatment of the thermodynamics to be obtained in the low temperature limit\cite{ma}, which was later proved to be asymptotically exact\cite{fisher}. The RG scheme showed how, in the limit of large disorder, the quantum system's behavior exhibits special characteristics that come about essentially because of enslavement of weak couplings by the strong ones, to a degree that is just not present in classical systems\cite{bhattphysica}. Such strong effects, which were further amenable to analytic RG treatments, rekindled the interest in rare fluctuation effects in quantum models with disorder.

It was recognized that the McCoy-Wu model was quite natural in the context of the one-dimensional random quantum Ising ferromagnet in a transverse magnetic field\cite{shankar}, the quantum mechanics entering through the non-commuting spin operators ($S_Z$ and $S_X$) along the Ising coupling (z) and the field (x) directions. The path integral representation of the quantum model led to a classical model in one higher dimension (i.e., two), and perfect correlations between the random bonds in one of the dimensions represented simply the time evolution in the path integral representation.  It was shown that approaching the ordered phase at zero temperature at low magnetic fields from the paramagnetic phase at high magnetic fields, the magnetic susceptibility diverged before the (quantum) phase transition, because of the effect of rare configurations of strong bonds. This generated a flurry of activity on transverse field quantum Ising models\cite{review}$^{,}$\cite{bhattspin}, including those with random bonds (e.g. the Ising spin glass in a transverse field). At the same time, it was recognized that similar effects were present in random, dimerized, quantum antiferromagnetic spin chains\cite{hyman}. 

The quantum spin glass models in $d=2$ and higher were not amenable to analytic study, but Monte Carlo simulations\cite{rieger} showed that the divergences of the non-linear (spin glass) susceptibility in the paramagnetic phase due to rare configurations continued to occur in $d = 2$ and $d = 3$ spatial dimensions. This strong effect in the thermodynamics of the quantum system, as opposed to rather weak effects in the classical system, can be physically understood by considering the path integral mapping of the quantum into a ($d+1$) dimensional classical system: while the classical random bond model has defects that are point-like in character, the quantum version has line defects (because of the perfect bond-correlation along the time direction).

While these studies showed quite convincingly the overall significant effect of rare configurations for the quantum Ising spin glass, it was after a very detailed numerical effort involving large amounts of computer time. Looking at details as a function of size or geometry of clusters which taken together give rise to the singular behavior was not deemed possible; this situation remains true today.

In the intervening years, there was not much further investigation of Lifshitz-tail\cite{abrahams}\cdash\cite{kramer} ideas in the context of the Anderson model\cite{anderson}; rather the excitement was concentrated mainly on the original issue raised by Anderson, namely, the extended-localized state (or the metal-insulator) transition. Starting with the pioneering numerical efforts of Kramer and MacKinnon\cite{mackinnon} and of McMillan and coworkers\cite{mcmillan}, the study of the metal-insulator transition for non-interacting electrons in a random potential with different symmetry properties (orthogonal, as in potential scattering, unitary as with a magnetic field; or symplectic – as with spin-orbit scattering), became a virtual industry, with the most complete work done by Slevin, Ohtsuki and coworkers \cite{slevin}\cdash\cite{asada}. So it was a complete surprise when our recent investigations\cite{johri} pointed out some rather remarkable, singular behavior in Anderson's original model, which had escaped attention for fifty years since it had been proposed. We elaborate upon our findings in the next section.

\section{Anderson's 1958 Model Revisited}
Anderson\cite{anderson} considered the one-electron Hamiltonian:
\begin{equation}
H = \sum_{i} \epsilon_i c_i^{\dagger} c_i  + \sum_{<ij>} V c_i^{\dagger} c_j,
\label{ham}
\end{equation}
where  i and j denote sites on a simple cubic lattice, $\epsilon_i$ are independent random variables drawn from a uniform probability distribution
\begin{equation}
P_U(\epsilon) = 1/W \mbox{ for } –W/2 < \epsilon < W/2 , \mbox{and zero otherwise},
\label{box}
\end{equation}
and sites $i$ and $j$ within the angular brackets $<ij>$ are nearest neighbors. We will generalize our study slightly, by considering some other probability distributions which are symmetric around zero (see below) and including dimensions $d = 1$, $2$ and $3$, but restricting our attention to $d$-dimensional hypercubic lattices. The model is characterized by the dimensionless disorder, $W/V$.

In three dimensions, for $W=0$, the density of states ($DOS$) is the familiar $DOS$ of the nearest neighbor tight-binding model on a simple cubic lattice extending from $E = -6V$ to $+6V$; needless to say, all electronic states are extended, being of the Bloch form. As $W/V$ is increased from zero, states near the band edges get localized, while those near the band center remain extended. Increasing $W/V$ further pushes the mobility edge [at $\pm E_c(W)$] separating the localized tail from the extended center closer to the band center at $E = 0$, until at $W \sim 16.5V$, i.e. for a disorder width of order the original bandwidth ($12V$), all states become localized [6]. In $d = 1$ and $2$, as discussed earlier, all states become localized for any nonzero $W/V$.

The other distributions we will discuss are Gaussian:
\begin{equation}
P_G(\epsilon) = (2\pi\sigma^2)^{-1/2} exp (-\epsilon^2/2\sigma^2),
\label{gaussian}
\end{equation}
where $\sigma=W/\sqrt{12}$ gives the same variance as $P_U$, the triangular distribution:
\begin{equation}
P_T(\epsilon) = 2(1- 2|\epsilon|/W)/W \mbox{ for } |\epsilon| < W/2, ), \mbox{0 otherwise},
\label{triangle}
\end{equation}
as well as a distribution with a power-law behavior at the edges ($\pm W/2$) and zero density of states at the origin:
\begin{equation}
P_{PL}(\epsilon) \propto |\epsilon|/(W^2 - 4\epsilon^2)^{\alpha-1}	\mbox{ for } |\epsilon|< W/2, \mbox{0 otherwise}
\label{power}
\end{equation}
In Eq. \ref{power} $\alpha > 0$, so that the distribution is integrable.

\begin{figure}[pb]
\vspace*{-10pt}
\centerline{\psfig{file=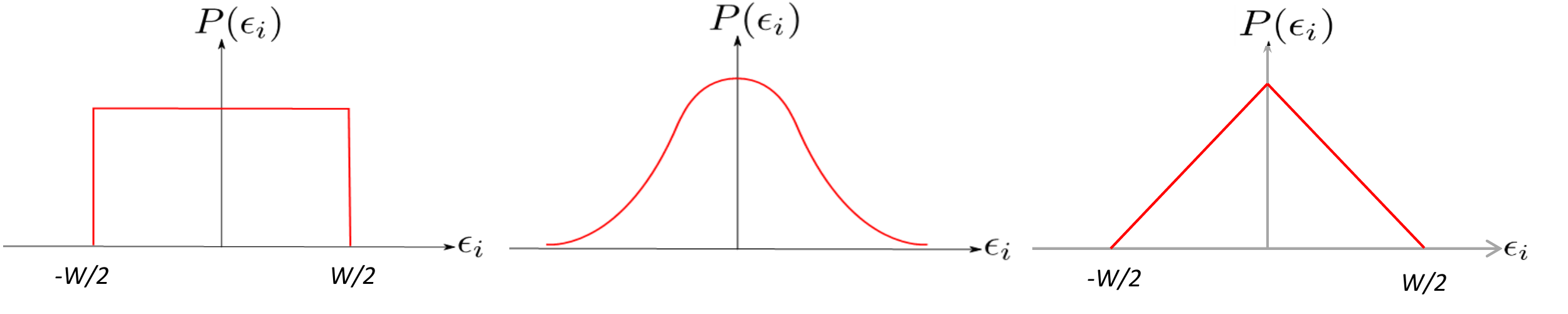,width=13cm}}
\vspace*{-10pt}
\caption{From left to right, uniform, Gaussian and triangular distribution of disorder. \label{dist}}
\end{figure}

Fig.~\ref{dist} depicts the first three of these distributions of the on-site energy, given by Eq. \ref{box}-\ref{triangle}. These distributions will help illustrate our findings and the conditions under which we get them.

In numerical studies of Anderson localization, two main quantities that have proved useful to calculate the localization length at any energy $E$, $\xi(E)$, have been the inverse participation ratio\cite{johri} $\langle IPR(E)\rangle$ , and the Lyapunov exponent\cite{kramer} $Ly(E)$ . The former is easily defined for any wavefunction $|\Psi> = \sum_i c_i |i>$, as

\begin{equation}
IPR_{\Psi} = (\sum_{i} c_i^4)/( \sum_i c_i^2)^2.
\label{iprpsi}
\end{equation}

Then
\begin{equation}
\langle IPR(E)\rangle = \langle IPR_{\Psi} \delta(E - E_{\Psi}) \rangle ,
\label{ipre}
\end{equation}
where the average is over states $\Psi$ that are at an energy in a small window around $E$, which can be made arbitrarily small in the thermodynamic limit, when the typical energy spacing scales as $1/L^d$, where $L$ is the linear dimension of the system. For a state that is localized on a length scale $\xi$, it is easily seen that $\langle IPR\rangle \sim \xi^{-d}$. For an extended state, on the other hand, $\langle IPR\rangle \sim L^{-d}$, where $L$ is the linear dimension of the system.

The Lyapunov exponent at energy $E$ is computed for systems of finite cross-section ($M^{d-1}$) and long length in one dimension, which is taken to the thermodynamic limit. It is defined in terms of the eigenvalues of the transfer matrix connecting the amplitudes of a wavefunction at position $i$ and ($i+1$) with those at ($i+1$) and ($i+2$).  In dimensions $d>1$, one has to evaluate it for different $M$ and take the limit as $M\rightarrow \infty$. For our purpose, it suffices to note that the limit, $Ly(E)$ characterizes the exponential decay of the wavefunction at long distances, at an energy $E$, and therefore,

\begin{equation}
Ly(E) \sim 1/\xi(E)
\label{ly}
\end{equation}

On the basis of (\ref{ipre}) and (\ref{ly}), one expects therefore that $\langle IPR(E)\rangle \sim [Ly(E)]^d$, and in particular, for $d=1$, they should be proportional to each other.

In the next section, we describe the results of our numerical investigations, by using $d = 1$, where states are localized at all energies for nonzero disorder ($W/V >0$). We discuss the situation in higher dimensions briefly in Section 4.

\section{Numerical Results (d=1)}

\begin{figure}[pb*]
\vspace*{-100pt}
\centerline{\psfig{file=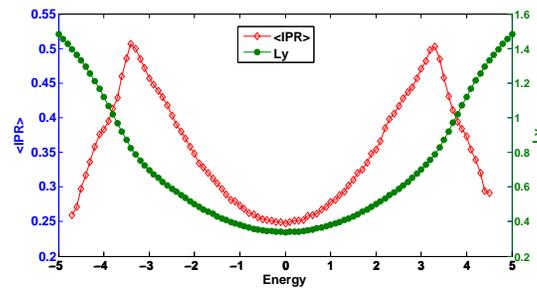,width=7.5cm}}
\vspace*{-95pt}
\caption{$\langle IPR(E)\rangle$ \& $Ly(E)$ for the uniform distribution with $W/V=6$. \label{ipr_ly}}
\end{figure}

Fig.~\ref{ipr_ly} shows $\langle IPR(E)\rangle$ and $Ly(E)$ as a function of $E$ for the Anderson Model (Eq. 1) with a uniform distribution (Eq. 2), in $d =1$ for $W/V = 6$. Since all states are localized, results converge rapidly as a function of size, and do not require extensive numerical effort. As can be seen, starting from the band center, both $\langle IPR(E)\rangle$ and $Ly(E)$ increase with $|E|$, indicating a decrease in the localization length with increasing $|E|$, as may be expected (band tails localize first even in $d = 3$). However, as one goes to larger $|E|$, $Ly(E)$ continues to rise, but $\langle IPR(E)\rangle$ goes through a maximum, and decreases, apparently going to zero at the band edge. This is consistent with Lifshitz's considerations ---the states at the absolute band edge for finite $W$ [given by $E_B = \pm(W/2 + 2dV)$ for hypercubic lattices in $d$-dimensions] are due to states on rare clusters of sites with on-site energies all at the edge of the disorder probability distribution, i.e. for $\epsilon_i = \pm W/2$ for all sites. Of course, such a configuration is exponentially rare in the size of the cluster $\sim exp [-c L^d]$ where $L$ is the linear dimension of the cluster, and has a vanishing probability in the thermodynamic limit. Of course, in such a case, the eigenstates are spread all over the cluster since there is no disorder, and would therefore have an $IPR \sim 1/L^d$. Particle in a box considerations suggest that the energy of these states would be $\sim 1/L^2$ away from the true band edge, implying that in the asymptotic region, $\langle IPR(E)\rangle \sim |E_{B}-E|^{d/2}$, which $\rightarrow 0$ as $E \rightarrow E_B$. Clearly in this (Lifshitz) regime, the extent of the wavefunction (measured by IPR) and the exponential decay length[]measured by $Ly(E)$] are not the same; one has \emph{two} length scales describing the electronic wavefunction. 

\begin{figure}[pb]
\vspace*{-100pt}
\centerline{\psfig{file=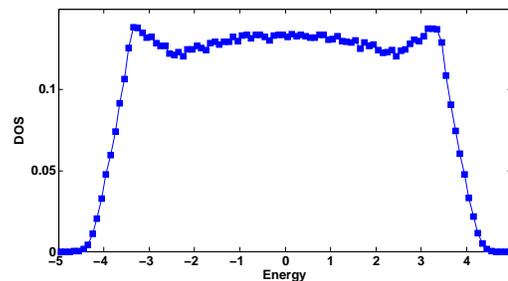,width=7.5cm}}
\vspace*{-95pt}
\caption{Density of states for the uniform distribution with $W/V=6$. \label{dos}}
\end{figure}

What is more striking than even the downturn is the manner in which the IPR changes course: there is a clear cusp in the data, implying non-analytic behavior. Furthermore, this non-analytic downturn happens at an energy $E_R$, where the density of states ($DOS$) is a maximum, not deep in the exponential tail of the $DOS$, as shown in Fig.~\ref{dos}. In fact, the $DOS$ also displays a cusp at the same energy. (This does not contradict the result of Edwards and Thouless\cite{edwards} who showed that the $DOS$ is analytic in an energy interval ($W/2-2dV$) around the band center; the singular behavior we see is outside this region). The cusp-like behavior is seen for intermediate and large ($W/V$); in $d=1$, we see it clearly for all $W/V \geq 3.8$. We remark that $Ly(E)$ does not show any sign of the singular behaviour at $E_R$.

As explained in Ref. ~\refcite{johri}, this singularity appears to be a consequence of the fact that typical Anderson localized states with a central site do not exist beyond $E_{R}$; the only states that are present beyond it are states that involve resonant tunneling between two or more nearest neighbor sites. In the Anderson model, for large disorder $W$, typical Anderson localized states have their energies perturbed by $O(V^2/W)$, whereas resonant states on two or more sites have energies shifted by $O(V)$; increasing the number of resonant sites only changes the prefactor of the $O(V)$ term, not the functional dependence. As a result, resonant states on sites near the disorder band edge “pull out” of the quasi-continuum of the typical Anderson-localized states, leaving behind non-analyticity in the density and nature of electronic states at the edge. 

$E_R$ thus appears to be the beginning of the Lifshitz tail, whose precise nature depends on short-distance rather than long-distance physics. Furthermore, since the smallest resonant cluster consists of 2 sites, its probability is not exponentially small, but only $O(V/W)$ for large $W$, so the number of resonant states beyond $E_{R}$ can be significant. For the case of the uniform distribution in $d = 1$, we find\cite{johri} that the maximum percentage of states beyond $E_R$ is $\approx 17\%$ (for $W/V \approx 3.8$). This fraction of tail states can hardly be called rare; hence the quotation marks in the title of this paper. We note in passing that we have shown [30] by explicit analytic calculation for a two-site Anderson model, that such cusp-like singularities exist in both $\langle IPR(E)\rangle$ and $DOS(E)$. 

\begin{figure}[pb]
\vspace*{-100pt}
\centerline{\psfig{file=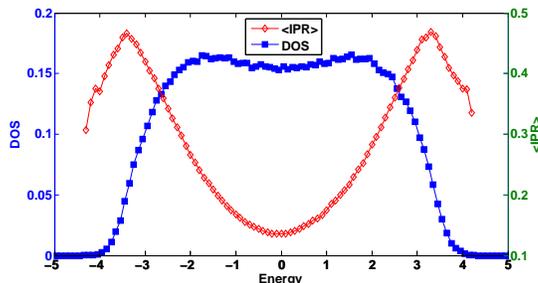,width=7.5cm}}
\vspace*{-95pt}
\caption{$DOS$ and $\langle IPR(E)\rangle$ for the triangular distribution of onsite energies with $W/V=6$. \label{triangle_dos}}
\end{figure}

The non-analytic behavior characterizing the separation of resonant states from typical Anderson localized states appears to exist for all forms of bounded disorder. Fig.~\ref{triangle_dos} shows $\langle IPR\rangle$ and $DOS$ as a function of $E$ for the triangular distribution (Eq. \ref{triangle}) for $W/V = 6$. Again the sharp maximum in $\langle IPR\rangle$ occurs at a $DOS$ that is not exponentially small, despite the smoother edge to the probability distribution of the disorder. On the other hand, for the Gaussian distribution of disorder (Eq. \ref{gaussian}), we find that both $Ly(E)$ and $\langle IPR(E)\rangle$ increase monotonically with $|E|$, with no sign of a turnaround or strong singularity for up to values of $|E|$ where the $DOS$ is down to $10^{-4}$ of its value at the band center. This is consistent with Wegner's result\cite{wegner} that the $DOS$ is analytic at all $E$ for Gaussian disorder.

For distributions like that in Eq. \ref{power} with several distinct edges ($|E| = W/2$, and $E = 0$), we find that the system develops more singular points dividing typical and resonant eigenstates. This is currently under further investigation.

\begin{figure}[pb]
\vspace*{-15pt}
\centerline{\psfig{file=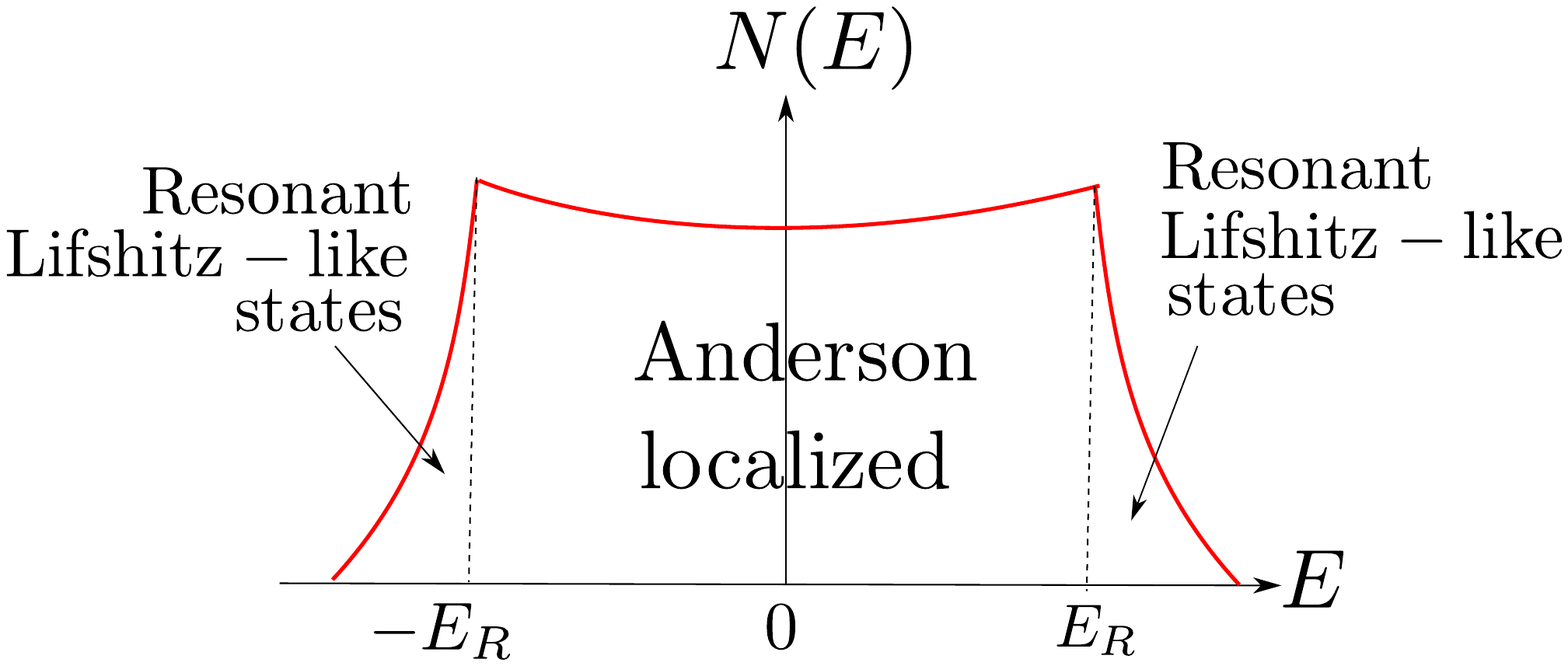,width=7.5cm}}
\vspace*{-210pt}
\caption{A schematic illustration showing the occurence of different kinds of wavefunctions in the insulating phase. \label{insulating}}
\end{figure}

On the basis of the above findings, we conclude that the Anderson model with bounded disorder in $d = 1$, when all states are localized, has a critical energy $E_R$ at least at large disorder, which demarcates the transition from typical Anderson localized states with a central site, to states that necessarily involve resonance between two or more nearest neighbor sites. This is depicted in Fig.~\ref{insulating}. When these resonant “Lifshitz-like” states occur at an energy that is forbidden to typical Anderson localized states, they maintain their strongly resonant character. On the other hand, when they occur at energies in the range allowed to typical Anderson localized states, they mix with the more abundant typical states, and lose their special resonant character. In that sense, this phenomenon appears to be analogous to the fact that localized states can only exist at energies where extended states do not; where they do, they mix and lose their localized character. Whether there is a further hierarchy among resonant states on different number of sites, and whether these are accompanied by (presumably weaker) singularities or not, remains a topic for further investigation.
 
\section{Higher Dimensions and Concluding Remarks}
In conclusion, we have shown that Anderson's original model of localization exhibits behavior that is richer than believed, so far. for the uniform distribution, as well as other bounded distributions of diagonal disorder, in the localized phase, the inverse participating ratio characterizing the extent of localized states exhibits a distinct, non-analytic behavior, which is to some extent reflected in the density of states as well. The energy $E_R$ at which this non-analyticity occurs separates typical Anderson localized states from states that have a resonant character on various length scales. 

This phenomenon, described above in Section 3 for $d = 1$ is not special to one dimension. As shown in Ref. ~\refcite{johri}, cusp-like singularities exist for large disorder in $d = 2$, and also in $d = 3$, in the insulating phase. Of course, detailed numbers are dimension dependent. It is interesting to speculate whether in $d = 3$, the existence of a second critical energy $E_C$ denoting the transition from extended to localized behavior will lead to an interplay of some sort. For example, would $|E_R|$ always be bigger than $|E_C|$ for all $W$, or could there be a transition from extended states directly into localized, resonant states as a function of energy for some value of the disorder? If so, would that be any different from the standard localization transition? In this regard, it is interesting to note that numerical investigations\cite{kramer}$^,$\cite{mackinnon}$^,$\cite{schreiber} of the localization transition in the Anderson model in $d = 3$ have found universal behavior for the transition at the band center ($E=0$) as a function of disorder ($W$), but have found non-universal results with similar size systems, when they study the transition at fixed $W$ as a function of energy (E). It is tempting to speculate that this may be connected to the change in the nature of the localized state as a function of energy at $E_R$, which is quite distinct from the true band edge $E_B$. Clearly, further research needs to be done to determine if the speculation has merit.

Another interesting result of our findings is that the original Anderson model\cite{anderson} provides a platform for an in-depth study of rare fluctuation physics. While analytic methods are most useful in the asymptotic exponential tail, where numerical methods are of little use, numerical methods can help provide detailed results due to resonant clusters of small finite sizes. If analytic methods can help decipher the nature of further (weaker) singularities due to the transitions from n-site clusters to ($n+1$)-site clusters that may exist as one descends down the tail of the distribution, there is the possibility that numerical methods may be able to pick them up quantitatively. This is because Anderson localization, being a single particle problem, requires time that grows as a power of the system size. Thus it is amenable to a much more intensive numerical investigation than many-body Hamiltonians with exponential growth of the Hilbert space with size. In addition, since most of the physics lies in the insulating phase, the sizes that will need to be considered will likely also be manageable.

\section*{Acknowledgments}
We thank Dr. Zlatko Papic for help with figures 1 and 5.



\begin{thebibliography}{0}    

\bibitem{anderson}P. W. Anderson, Physical Review {\bf 109}, 1492 (1958).
\bibitem{mott}N. F. Mott Jour. Non-Crystalline Solids, {\bf 1}, 1 (1968).
\bibitem{motttwose}N. F. Mott and W. D. Twose, Adv. Phys. {\bf 10}, 107 (1961).
\bibitem{abrahams}E. Abrahams, P. W. Anderson, D. C. Licciardello and T. V. Ramakrishnan, Physical Review Letters {\bf 42}, 673 (1979).
\bibitem{lifshitz}I. M. Lifshitz, Adv. Phys. {\bf 13}, 483 (1964); Soviet Physics Uspekhi {\bf 7}, 549 (1965).
\bibitem{kramer}B. Kramer and A. MacKinnon, Rep. Prog. Phys. {\bf 56}, 1469 (1993).
\bibitem{halperin}B. I. Halperin and Melvin Lax, Physical Review {\bf 148}, 722 (1966); ibid. {\bf 153}, 802 (1967).
\bibitem{mccoywu}B. M. McCoy and T. T. Wu, Physical Review {\bf 176}, B631 (1968); ibid. {\bf 188}, 982 (1969).
\bibitem{mccoy}B. M. McCoy, Physical Review Letters {\bf 23}, 383 (1969); Physical Review {\bf 188}, 1014 (1969).
\bibitem{griffiths}R. B. Griffiths, Physical Review Letters {\bf 23}, 17 (1969).
\bibitem{dhar}D. Dhar and M. Barma, Jour. Stat. Phys. {\bf 22}, 259 (1980); D. Dhar in Stochastic Processes: Formalism and Applications, G. S. Aggarwal and S. Dattagupta, editors (Springer, Berlin, 1983).
\bibitem{randeria}M. Randeria, J. P. Sethna, and R. G. Palmer, Physical Review Letters {\bf 54}, 1321 (1985).
\bibitem{ogielski}A. Ogielski, Physical Review B {\bf 32}, 7384 (1985).
\bibitem{svedlindh} P.Svedlindh, P.Granberg, P. Nordblad, L. Lundgren and H. S. Chen, Physical Review B {\bf 35}, 268 (1987); K. Gunnarsson, P. Svedlindh, P. Nordblad, L. Lundgren, H. Hruga and A. Ito, Physical Review Letters {\bf 61}, 754 (1988).
\bibitem{bhattrice}R. N. Bhatt and T. M. Rice, Philos. Mag. {\bf 42B}, 859 (1980); M. Rosso, Physical Review Letters {\bf 44}, 1541 (1980); R. N. Bhatt, Physical Review Letters 48, 707 (1982).
\bibitem{bhattlee}R. N. Bhatt and P. A. Lee, Bull. Am. Phys. Soc., {\bf 25}, 206 (1980); Journal of Applied Physics {\bf 52}, 1703 (1981); Physical Review Letters {\bf 48}, 344 (1982).
\bibitem{ma}S. K. Ma, C. Dasgupta and C. K. Hu, Physical Review Letters {\bf 43}, 1899 (1979); C. Dasgupta and S. K. Ma, Physical Review B {\bf 22}, 1305 (1980)
\bibitem{fisher}D S Fisher, Physical Review B {\bf 50}, 3799 (1994).
\bibitem{bhattphysica}R.N. Bhatt, Physica {\bf 109-110} B\&C, 2145 (1982)
\bibitem{shankar}R. Shankar and G. Murthy, Physical Review B {\bf 36}, 536 (1987); D. S. Fisher, Physical Review Letters {\bf 69}, 534 (1992); Physical Review B {\bf 51}, 6411 (1995). 
\bibitem{review}For a review, see Quantum Ising Phases and Transitions in Transverse Ising Models, B. K. Chakrabarti, A. Dutta and P.Sen (Springer, Berlin, 1996).
\bibitem{bhattspin} R. N. Bhatt, in Spin Glasses and Random Fields, ed. A. P. Young (World Scientific, 1997).
\bibitem{hyman} R.A. Hyman, K. Yang, R.N. Bhatt and S.M. Girvin, Physical Review Letters {\bf 76}, 839 (1996)
\bibitem{rieger}H. Rieger and A. P. Young, Physical Review B {\bf 54}, 3328 (1996); Muyu Guo, R. N. Bhatt and David A. Huse, Physical Review B {\bf 54}, 3336 (1996).
\bibitem{mackinnon}A. MacKinnon and B. Kramer, Physical Review Letters {\bf 47}, 1546 (1981).
\bibitem{mcmillan}W. L. McMillan, Physical Review B {\bf 31}, 344 (1985); Avinash Singh and W. L. McMillan, Jour. of Physics C {\bf 17}, 2097 (1985).
\bibitem{slevin}K. Slevin and T. Ohtsuki, Physical Review Letters {\bf 78}, 4083 (1997).
\bibitem{asada}Y. Asada, K. Slevin, and T. Ohtsuki. Physical Review Letters {\bf 89}, 256601 (2002).
\bibitem{johri}S. Johri and R. N. Bhatt, arXiv1106.1131 (submitted for publication).
\bibitem{evers}F. Evers and A. D. Mirlin, Reviews of Modern Physics {\bf 80}, 1355 (2008).
\bibitem{edwards}J. T. Edwards and D. J. Thouless, Journal of Physics C {\bf 4}, 453 (1971).
\bibitem{johriprep}S. Johri and R. N. Bhatt (in preparation)
\bibitem{wegner}F. Wegner, Z. Phys. B {\bf 44}, 9 (1981).
\bibitem{schreiber}B. Kramer, K. Broderix, A. MacKinnon and M. Schreiber, Physica A {\bf 167}, 163 (1990).
\end{thebibliography}
\end{document}